\documentclass[fleqn,twoside]{article}
\usepackage{espcrc2}
\usepackage{graphicx}
\newcommand{\be}{\begin{eqnarray}}
\newcommand{\ee}{\end{eqnarray}}
\newcommand{\ba}{\begin{array}}
\newcommand{\ea}{\end{array}}

\newcommand{\dsla}{D \hspace{-7pt}/~}

\newcommand{\partsla}{\partial \hspace{-7pt}/~}

\newcommand{\vsa}{\rlap{\hbox{$\longrightarrow$}}
                   \raise 7pt \hbox{\scriptsize VSA~~}}
\newcommand{\ope}{\rlap{\hbox{$\Longrightarrow$}}
                   \raise 4.2ex \hbox{\hspace{-0.3ex}\scriptsize OPE~~}}

\newcommand{\nabr}{\rlap{\hbox{$\nabla$}}
                   \raise 8 pt \hbox{$\hspace{-0.05cm}\leftarrow$}}
\newcommand{\nabd}{\rlap{\hbox{$D$}}
                   \raise 7 pt \hbox{$\hspace{-0.05cm}\leftarrow$}}
\newcommand{\rvD}{\rlap{\hbox{$\vec{D}$}}
                   \raise 8 pt \hbox{$\hspace{-0.05cm}\leftarrow$}}
\newcommand{\dslar}{\rlap{\hbox{$\dsla$}}
                   \raise 8 pt \hbox{$\hspace{-0.05cm}\rightarrow$}}
\newcommand{\dslal}{\rlap{\hbox{$\dsla$}}
                   \raise 8 pt \hbox{$\hspace{-0.05cm}\leftarrow$}}
\newcommand{\dl}{\rlap{\hbox{$D_\mu$}}
                   \raise 8 pt \hbox{$\hspace{-0.05cm}\leftarrow$}}
\newcommand{\partr}{\rlap{\hbox{$\partial_\mu$}}
                   \raise 8 pt \hbox{$\hspace{-0.05cm}\reftarrow$}}
\newcommand{\partl}{\rlap{\hbox{$\partial_\mu$}}
                   \raise 8 pt \hbox{$\hspace{-0.05cm}\leftarrow$}}
\newcommand{\partslar}{\rlap{\hbox{$\partsla$}}
                   \raise 8 pt \hbox{$\hspace{-0.05cm}\rightarrow$}}
\newcommand{\partslal}{\rlap{\hbox{$\partsla$}}
                   \raise 8 pt \hbox{$\hspace{-0.05cm}\leftarrow$}}
\newcommand{\lw}[1]{\smash{\lower1.ex\hbox{#1}}}

\newcommand{\Tsukuba}
{Institute of Physics, University of Tsukuba, Tsukuba,
 Ibaraki 305-8571, Japan}
\newcommand{\CCP}
{Center for Computational Physics, University of Tsukuba,
 Tsukuba, Ibaraki 305-8577, Japan}
\newcommand{\ICRR}
{Institute for Cosmic Ray Research, University of Tokyo,
 Kashiwa, Chiba 277-8582, Japan}
\newcommand{\KEK}
{High Energy Accelerator Research Organization(KEK), Tsukuba,
 Ibaraki 305-0801, Japan}
\newcommand{\YITP}
{Yukawa Institute for Theoretical Physics, Kyoto University,
 Kyoto 606-8502, Japan}
\title{$B$ meson $B$-parameters and the decay constant
       in two-flavor dynamical QCD \thanks{Talk presented by N. Yamada.}}
\author{
  JLQCD Collaboration:
  N.~Yamada\address{\KEK\\[-1ex]},
  S.~Aoki\address{\Tsukuba\\[-1ex]},
  R.~Bulkhalter$^{\rm b,}$\address{\CCP\\[-1ex]},
  M.~Fukugita\address{\ICRR\\[-1ex]},
  S.~Hashimoto$^{\rm a}$,
  K-I.~Ishikawa$^{\rm a}$,
  N.~Ishizuka$^{\rm b,c}$,
  Y.~Iwasaki$^{\rm b,c}$,
  K.~Kanaya$^{\rm b,c}$,
  T.~Kaneko$^{\rm a}$,
  Y.~Kuramashi$^{\rm a}$,
  M.~Okawa$^{\rm a}$,
  T.~Onogi\address{\YITP\\[-1ex]},
  S.~Tominaga$^{\rm c}$,
  N.~Tsutsui$^{\rm a}$,
  A.~Ukawa$^{\rm b,c}$,
  T.~Yoshi\'e$^{\rm b,c}$
  }
\begin{document}
\begin{abstract}
  We present a two-flavor dynamical QCD calculation of the $B$ meson 
  $B$ parameters and decay constant.
  We use NRQCD for heavy quark and the nonperturbatively
  $O(a)$-improved Wilson action for light quark at $\beta=5.2$
  on a $20^3\times 48$ lattice.
  We confirm that the sea quark effect increases the heavy-light
  decay constant, while estimate of its magnitude depends significantly
  on the fitting form in the chiral extrapolation.
  For the $B$ parameters, on the other hand, we do not find a
  significant sea quark effect.
  The chiral extrapolation with logarithmic term is examined for both
  quantities and compared with the prediction of ChPT.
\vspace{-120mm}
\begin{flushright}
\large KEK-CP-113
\end{flushright}
\vspace{105mm}
\end{abstract}
\maketitle
\section{Introduction}
In recent lattice calculations, there have been several indications 
that the presence of sea quark increases the $B$ meson decay constant
$f_B$ significantly, \textit{i.e.} by about 10--15\% 
\cite{Bernard:2001ki}.
Then, the study of the sea quark effect on the $B$ parameters has
a great phenomenological importance, because the relevant physical
quantity to the $B-\bar{B}$ mixing is the combination $f_B^2 B_B$.

In this work we perform unquenched lattice calculations of both $f_B$
and $B_B$ using a consistent formulation.
We use the nonperturbatively $O(a)$-improved Wilson quark action for
both sea and valence light quarks, and the NRQCD action for heavy
quark.
Using the same dynamical QCD configurations, a search for sea quark
effects in the light hadron spectrum is also being performed
\cite{Hashimoto_lat2001}.
Quenched calculations of $f_B$ and $B_B$ using the NRQCD action 
have been presented previously \cite{Ishikawa:2000xu,Yamada:2001ym}. 

In full QCD, the effect of virtual pion loops leads to chiral
logarithms, a specific dependence of physical quantities on the sea
quark mass.
Hence, it provides a non-trivial test of the sea quark effect in
lattice simulations.
For $f_B$ and $B_B$ chiral perturbation theory (ChPT) is
extensively studied in \cite{Booth:1995hx,Sharpe:1996qp} including its
extension to the quenched and the partially quenched cases, where sea
and valence quark masses can be different.
We test the chiral logarithm using our numerical data.

\section{Simulation details}

We generate a set of unquenched gauge configurations with the
plaquette gauge action and two-flavors of nonperturbatively
$O(a)$-improved Wilson fermions \cite{Jansen:1998mx}. 
Simulations are performed on a $20^3\times 48$ lattice at $\beta$=5.2,
which roughly corresponds to the lattice spacing $a\sim$ 0.1~fm.
We take five different sea quark masses covering
$m_{PS}/m_V$=0.6--0.8.
For comparison we also perform a quenched simulation at $\beta$=6.0 on
the same lattice size.
Further details of our simulations are given in
\cite{Hashimoto_lat2001}.

We use the NRQCD action for heavy quark including all $1/m_Q$
corrections consistently in the action and operators.
In the calculations of $f_B$ and $B_B$ we follow the methods adopted
in the previous works \cite{Ishikawa:2000xu,Yamada:2001ym}. 
In particular, the $B$ parameters are computed with four different
prescriptions, all of which have equivalent accuracy up to higher order
contributions we neglect.
We may estimate the systematic uncertainty taking the differences
among them \cite{Yamada:2001ym}.

\section{Decay constant}

\begin{figure}
  \leavevmode
  \includegraphics*[width=7.5cm,clip]{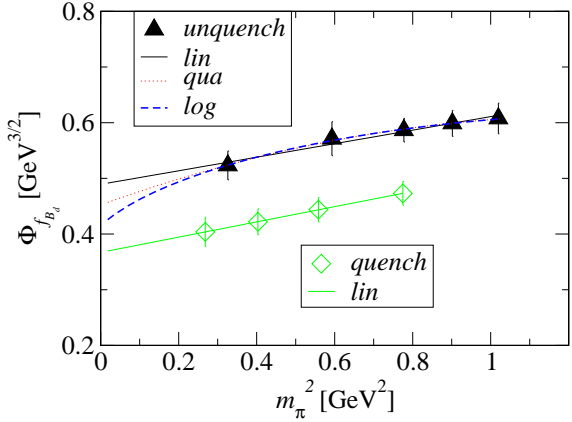}
  \vspace{-6ex}
  \caption{Chiral extrapolation of $\Phi_{f_{B_d}}$.}
  \label{fig:chi_phi_fB}
  \vspace{-2ex}
\end{figure}

Figure~\ref{fig:chi_phi_fB} shows the chiral behavior of the heavy-light
decay constant $\Phi_{f_B}\simeq f_B\sqrt{m_B}$ with $m_Q\sim m_b$.
Both horizontal and vertical axes are normalized with a proper power
of $r_0$ and converted to the physical unit using the lattice result
for $r_0 m_\rho$ combined with physical input for $m_\rho$.
If we compare the unquenched result (filled triangles) with the
quenched data (open diamonds), we clearly find the increase of about
15\% with the unquenching as found previously \cite{Bernard:2001ki}

We consider the following three forms for the chiral extrapolation 
\begin{eqnarray}
  \begin{array}{l@{\;\;:\;\; }l}
    \mbox{``lin''}& a_{0}\, (1 + a_{1}X ),\\
    \mbox{``qua''}& b_{0}\, (1 + b_{1}X + b_{2}X^2 ),\\
    \mbox{``log''}& c_{0}\, (1 + c_{1}X + c_{2}X\ln X ),
  \end{array}
  \label{eq:chi_phi_r}
\end{eqnarray}
with $X=r_0^2m_\pi^2$.
The ``log'' form is motivated by the ChPT result for heavy-light mesons
\cite{Booth:1995hx,Sharpe:1996qp}.
For two-flavor QCD, it predicts 
\begin{equation}
  r_0^2 c_2 = - \frac{3}{4} \frac{1+3 g^2}{(4\pi f)^2},
\end{equation}
with $f$ the pion decay constant.
The coupling $g$ describes $B^*B\pi$ interaction, which may be
evaluated from $D^*$ decays using heavy quark symmetry.
If we take a range $g$=0.2--0.7 rather conservatively covering
$D^*\rightarrow D\pi$ decay \cite{Stewart:1998ke} and the recent $D^*$ 
width experiment \cite{Anastassov:2001cw}, then we obtain 
$r_0^2c_2$=$-$0.5(2)~GeV$^{-2}$.

We find that both ``qua'' and ``log'' fits describe our unquenched
data well, while the linear fit gives somewhat poorer $\chi^2$ than
the other two.
For the ``log'' fit we obtain $r_0^2c_2$=$-$0.4(3)~GeV$^{-2}$, which is
consistent with the above phenomenological estimate.

On the other hand, in quenched QCD, no deviation from linearity is
observed in our data, and the three fit forms lead to almost identical
result in the chiral limit.
This behavior contradicts the prediction of quenched ChPT (QChPT) 
\cite{Booth:1995hx,Sharpe:1996qp}, which gives even stronger
logarithmic dependence.
There are, however, many unknown parameters in QChPT, so it
is still possible to find a parameter region which mimics the linear
behavior in our light quark mass regime.

\begin{figure}
  \leavevmode
  \includegraphics*[width=7.5cm,clip]{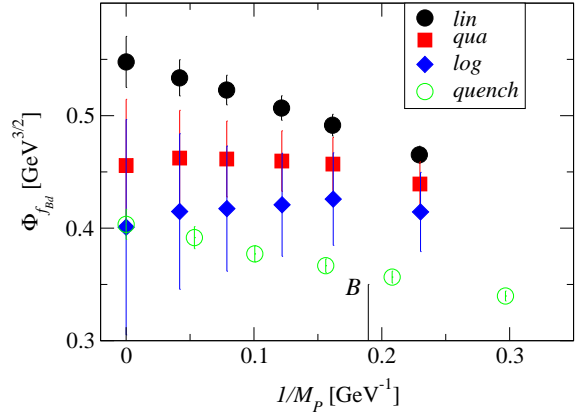}
  \vspace{-6ex}
  \caption{The $1/M_P$ dependence of $\Phi_{f_{B_d}}$.}
  \vspace{-2ex}
  \label{fig:mas_phi_fB}
\end{figure}

Since we are not able to decide the fit form with our current
unquenched data, the uncertainty in the chiral limit is quite large. 
Figure~\ref{fig:mas_phi_fB} shows the $1/M_P$ dependence of
$\Phi_{f_{B_d}}$ at the physical light quark mass.
At the physical $B$ meson mass, we observe that the two-flavor result 
(filled symbols) is 14--25\% larger than the quenched result (open
circle), depending on the fit form.

\section{$B$ parameters}

We also examine the chiral behavior of $B_B$ by applying the fit forms
in (\ref{eq:chi_phi_r}), but, because the light quark mass dependence
itself is quite small, we do not observe any significant dependence on
the fit functions.
A weak dependence on sea quark mass is also suggested by ChPT, which gives 
\begin{equation}
  r_0^2c_2 = - \frac{3}{2} \frac{1-3 g^2}{(4\pi f)^2}
  = -0.1(4)~\mbox{GeV}^{-2}
\end{equation}
for $B_B$.

\begin{figure}
  \vspace*{-1ex}
  \leavevmode
  \includegraphics*[height=7.5cm,clip,angle=270]{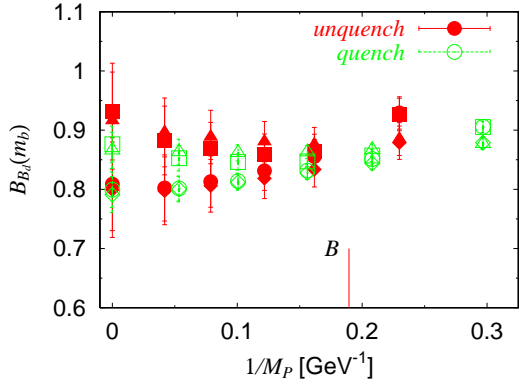}
  \vspace{-6ex}
  \caption{The $1/M_P$ dependence of $B_{B_d}(m_b)$.}
  \label{fig:mas_BB}
  \vspace{-2ex}
\end{figure}

Figure~\ref{fig:mas_BB} shows the $1/M_P$ dependence of $B_B(m_b)$ for
both quenched (open symbols) and unquenched QCD (filled symbols).
Four different symbols correspond to different prescriptions defined
in \cite{Yamada:2001ym}.
Within the uncertainty indicated by the different prescriptions, we
find no significant effect of unquenching.
$B_{S_s}(m_b)$ is also analyzed in a similar manner and
we arrive at the same conclusion.

\section{Conclusions}

Our preliminary results in two-flavor QCD are summarized as follows, 
\begin{eqnarray}
  f_{B_d}         &=& 190(14)(07)(19)\, \mbox{MeV},\nonumber\\
  f_{B_s}/f_{B_d} &=& 1.184(26)(20)(15), \nonumber\\
  B_{B_d}(m_b)    &=& 0.872(39)(04)(73),\nonumber\\
  B_{B_s}/B_{B_d} &=& 0.999(12)(04),\nonumber\\
  B_{S_s}(m_b)    &=& 0.858(33)(07)(72),\nonumber
\end{eqnarray}
where errors are from statistics, chiral extrapolation, and
systematics.
We discard the linear fit from the final results because of its poor
$\chi^2$ value. 
The central value is obtained by taking an average over 
``qua'' and ``log'' results, and their difference is taken as a
systematic error from the chiral extrapolation. 
Some of the systematic errors in the $B$ parameters are estimated by
the scatter with different prescriptions as in \cite{Yamada:2001ym}.
Other uncertainties are estimated by a naive dimensional counting,
assuming $\Lambda_{\rm QCD}$=500~MeV and
$\alpha_s$=$\alpha_V(1/a)$=0.267. 

Our data for $f_B$ and $B_B$ are consistent with the prediction of ChPT.
This is different from the light hadron sector where we find that
ChPT fails to reproduce the lattice data for pion mass and decay
constant \cite{Hashimoto_lat2001}.
To investigate these issues we are currently accumulating further
statistics, which also helps to reduce the statistical error in our
results.

\vspace*{5mm}
This work is supported by the Supercomputer Project No.66 (FY2001)
of High Energy Accelerator Research Organization (KEK),
and also in part by the Grant-in-Aid of the Ministry of Education
(Nos. 10640246, 11640294, 12014202, 12640253, 12640279, 12740133,
13640260 and 13740169).
K-I.I and N.Y are supported by the JSPS Research Fellowship.

\end{document}